\documentclass[twocolumn,preprintnumbers,amsmath,amssymb,floatfix,superscriptaddress,nofootinbib]{revtex4}

\usepackage{amsmath,hyperref,amssymb}
\usepackage{graphicx}
\usepackage{dcolumn}
\usepackage{bm}
\usepackage{verbatim}
\usepackage{tabularx}
\usepackage{slashed}
\usepackage{cancel}
\usepackage[boxsize=2em,aligntableaux=center]{ytableau}
\usepackage{float}
\usepackage{color}
\usepackage{url}
\usepackage[utf8]{inputenc}
\usepackage{enumerate}
\usepackage{array}

\newcolumntype{x}[1]{>{\centering\arraybackslash\hspace{0pt}}p{#1}}
\def\be{\begin{equation}}
\def\ee{\end{equation}}
\def\bea{\begin{eqnarray}}
\def\eea{\end{eqnarray}}

\newcommand{\gev}{{\rm GeV}}

\begin{document}


\title{Muons in supernovae: implications for the axion-muon coupling}

\author{Robert Bollig}
\affiliation{Max-Planck-Institut f\"ur Astrophysik\\
Karl-Schwarzschild-Stra\ss e 1, 85748 Garching, Germany}
\author{William DeRocco}
\affiliation{Stanford Institute for Theoretical Physics, \\
Stanford University, Stanford, CA 94305, USA}
\author{Peter W.~Graham}
\affiliation{Stanford Institute for Theoretical Physics, \\
Stanford University, Stanford, CA 94305, USA}
\author{Hans-Thomas Janka}
\affiliation{Max-Planck-Institut f\"ur Astrophysik\\
Karl-Schwarzschild-Stra\ss e 1, 85748 Garching, Germany}

\vspace*{1cm}

\begin{abstract} 

The high temperature and electron degeneracy attained during a supernova allow for the formation of a large muon abundance within the core of the resulting proto-neutron star. If new pseudoscalar degrees of freedom have large couplings to the muon, they can be produced by this muon abundance and contribute to the cooling of the star. By generating the largest collection of supernova simulations with muons to date, we show that observations of the cooling rate of SN 1987A place strong constraints on the coupling of axion-like particles to muons, limiting the coupling to $g_{a\mu} < 10^{-7.5}~\text{GeV}^{-1}$.

\end{abstract}

\maketitle

\textit{\textbf{Note on this version:} This arXiv posting presents a consolidated version of our paper that, relative to the published one, incorporates both the modifications discussed in the published Erratum as well as a variety of small corrections and improvements in the overall presentation. We are extremely grateful to Georg Raffelt for pointing out these issues to us.}

\section{Introduction}

Axions are hypothetical light pseudoscalar degrees of freedom. Initially proposed to solve the strong CP problem~\cite{Peccei:1977hh,Peccei:1977ur,Weinberg:1977ma}, they are a generic prediction of string theory~\cite{Arvanitaki:2009fg} and a good candidate for dark matter~\cite{Preskill:1982cy,Abbott:1982af,Dine:1982ah,Duffy:2009ig,Ringwald:2016yge}. As a result, a large research effort is dedicated to searching for these ``axion-like particles'' (henceforth axions), mainly through their coupling to photons~\cite{Anastassopoulos:2017ftl,PhysRevLett.120.151301,Dine:1982ah,Brubaker:2016ktl,Betz:2013dza}, electrons~\cite{Barth_2013}, or nucleons~\cite{JacksonKimball:2017elr} (see~\cite{Graham:2015ouw} for a review). The coupling to muons, however, has not been well studied, partially by virtue of the fact that the high muon rest mass and short lifetime make performing precision muon experiments over long timescales difficult. Luckily, astrophysics provides us with an alternate means to probe this coupling: the cooling of supernovae.\footnote{Neutron star cooling may also place a bound, but rough estimates indicate that the supernova bound is stronger. This is a future avenue to explore.}

As noted recently~\cite{Lohs:2015qyn,Brust:2013xpv,Bollig:2017lki}, though the temperature of a supernova hardly rises above $\sim 50 \, \text{MeV}$, there is still a large proportion of thermal photons and neutrinos with energies well above the muon rest mass. Additionally, the electrons acquire a large chemical potential allowing their conversion to muons, with the production of muon antineutrinos compensating to maintain a net zero muon number. Due to the small difference in the neutral current cross-section between $\nu$ and $\bar{\nu}$, muon antineutrinos diffuse out of the PNS at a higher rate than muon neutrinos, leading to the accumulation of a net muon number within the PNS in a process known as ``muonization.'' Recent work has shown this effect may play a non-negligible role in facilitating supernova explosions~\cite{Bollig:2017lki}.

The resulting muon abundance could produce a large axion flux. Here, we exclusively consider axions with mass $m_a \ll $ MeV and muon-dominated interactions, coupled via\footnote{Note that while these couplings are equivalent for the tree-level processes discussed in this paper, they need not be equivalent in general.}
\be
\mathcal{L} \subset g_{a\mu}(\partial_{\sigma} a)\bar{\psi}_{\mu}\gamma^{\sigma}\gamma_5\psi_{\mu} \equiv -i g_{a\mu}(2m_{\mu})\bar{\psi}_{\mu}\gamma_5\psi_{\mu}a \,,
\ee
where $m_{\mu}$ is the muon mass, $a$ the axion field, $\psi_{\mu}$ the muon spinor, and $g_{a\mu}$ the axion-muon coupling. Bounds on $g_{a\mu}$ arise from axion contributions to the muon $g - 2$, at the level of $g_{a\mu} \lesssim 10^{-2.4}~\text{GeV}^{-1}$~\cite{Andreas:2010ms}, as well as from cosmology~\cite{Brust:2013xpv,Baumann:2016wac,DEramo:2018vss} (see Supplemental Material for a detailed discussion of these limits). The bounds we calculate here extend the experimentally-excluded region by over five orders of magnitude.

Observations of SN 1987A neutrinos indicate that the resulting proto-neutron star (PNS) cooled in roughly ten seconds. A new particle transferring energy more efficiently than the neutrinos would shorten this timescale, leading to the oft-cited ``cooling bound''~\cite{Raffelt:1996wa,PhysRevLett.60.1797,PhysRevD.39.1020,RAFFELT19901,Chang:2018rso,DeRocco:2019jti,Carenza_2019} on the axion luminosity of $L_a \lesssim 3\times10^{52}$ erg/s.\footnote{Ref.~\cite{Bar+2019} questions this bound, but is based on a speculative thermonuclear explosion scenario~\cite{BlumKushnir2016,KushnirKatz2015} that conflicts with stellar evolution theory, supernova nucleosynthesis, and the existence of a neutron star in SN~1987A, which is strongly suggested by interpretations of recent observations by the Atacama Large Millimeter Array~\cite{Cigan+2019,Page+2020}.}

Note that a rough estimate of this bound was made in Ref.~\cite{Brust:2013xpv}, but due to its highly approximative nature, the result was subject to a large degree of uncertainty~\cite{DEramo:2018vss}. This is not unexpected, as the high muon rest mass means both the muon density and reaction rates depend sensitively on the core temperature (and muon and electron chemical potentials) of SN 1987A. In this paper, we significantly improve on the previous estimate by running dedicated simulations that make use of recent results~\cite{Fischer+2014,Oertel+2017,Fischer+2017,Demorest+2010,Antoniadis+2013,Cromartie+2020,Abbott+2018,Bauswein+2017,Essick:2020flb} constraining the PNS equation of state and mass, allowing us to reduce the uncertainty and set a robust bound on the coupling over five orders of magnitude. Interestingly, these new constraints on the EoS and mass lead to a generically higher core temperature, which suggests that existing bounds on other axion couplings may be strengthened by including this microphysics as well.

\section{Profiles from simulations}

As with any supernova bound, it is critical to assess the robustness of our results to variations in the choice of model. To do this, we ran four simulations that span a range of allowable final neutron star (NS) masses for SN1987A (see Table~\ref{table:params}). The simulations were performed in spherical symmetry (1D) with the \textsc{Prometheus-Vertex} code with general-relativistic corrections and six-species neutrino transport, solving iteratively the two-moment equations for neutrino energy and momentum with a Boltzmann closure~\cite{RamppJanka2002} and using the full set of neutrino processes listed in~\cite{Janka2012} and~\cite{Bollig:2017lki}. PNS convection was taken into account by a mixing-length treatment and explosions were artificially triggered a few 100\,ms after bounce at the progenitor's Fe/Si or Si/O composition interface as described in~\cite{Mirizzi+2016}.

\begin{table*}
\begin{center}
\begin{tabular}{|c|c|c|c|c|}
\hline
Model name & Equation of state & Progenitor mass ($M_{\odot}$) & NS bary. mass ($M_{\odot}$) & NS grav.\ mass ($M_{\odot}$) \\ 
\hline
SFHo-18.8 & SFHo~\cite{Steiner:2012rk}    &18.8~\cite{Sukhbold+2018}    & 1.351 & 1.241 \\ 
SFHo-18.6 & SFHo~\cite{Steiner:2012rk}    &18.6~\cite{Woosley+2002}     & 1.553 & 1.406 \\ 
SFHo-20.0 & SFHo~\cite{Steiner:2012rk}    &20.0~\cite{WoosleyHeger2007} & 1.947 & 1.712 \\ 
LS220-20.0 & LS220~\cite{Lattimer:1991nc} &20.0~\cite{WoosleyHeger2007} & 1.926 & 1.707 \\ 
\hline
\end{tabular}
\end{center}
\caption{Supernova model parameters and resulting NS baryonic and gravitational masses. We place our final bound with SFHo-18.8, as it produces the weakest constraint.}
\label{table:params}
\end{table*}

The main astrophysical uncertainties are connected to
the neutron star mass in SN~1987A and the equation of
state at supernuclear densities. Model SFHo-18.6 has 
a canonical neutron star mass well within the range 
expected for the compact remnant in SN~1987A, while 
in models SFHo-20.0 and LS220-20.0 the neutron star
masses are near the upper edge of the expected 
range, and in model SFHo-18.8 the neutron star mass is at the lower edge of the allowed range~\cite{Page+2020}. The SFHo equation of state
is fully compatible with all current constraints from
nuclear theory and experiment~\cite{Fischer+2014,Oertel+2017,Fischer+2017}
and astrophysics, including pulsar mass
measurements~\cite{Demorest+2010,Antoniadis+2013,Cromartie+2020}
and the radius constraints deduced
from gravitational-wave and Neutron Star Interior Composition Explorer measurements~\cite{Abbott+2018,Bauswein+2017,Essick:2020flb}.
Results for the long-used LS220 equation of state are shown 
for reference and comparison. Note that these equations of state are
considerably softer than those of previous works on
axion emission from supernovae~\cite{Chang:2016ntp,Fischer:2016cyd,Nakazato:2012qf},
which employed stiff equations of state that are increasingly disfavored by the constraints above.
The adoption of softer equations of state generically results in a smaller radius and higher 
temperature than in PNS models with stiff equations of state. It should be noted that though this change is most relevant for the axion-muon coupling, the generically higher temperatures will influence many existing supernova limits on new particles, and these results should be revisited with the new EoS constraints in mind.

\begin{figure}
  \centering
  \includegraphics[width=0.45\textwidth]{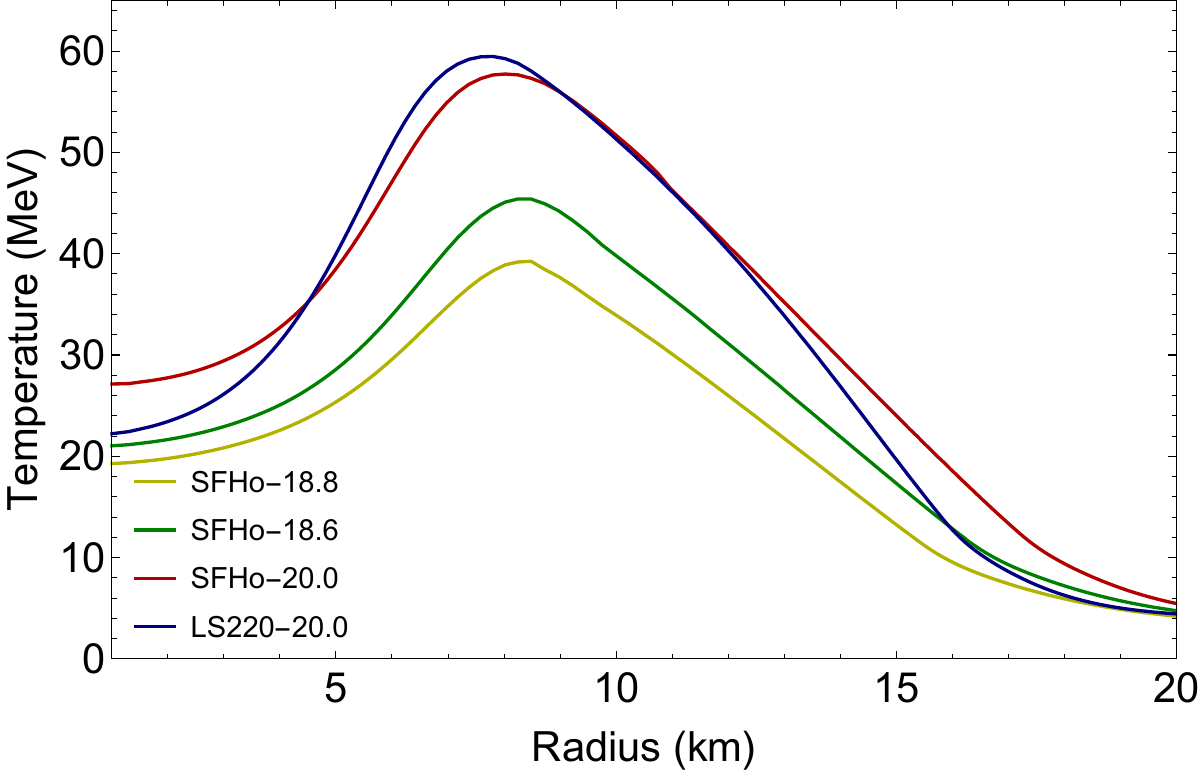}
  \caption{Temperature profile for various models at 1\,s post-bounce.
     \label{Fig: temp}}
\end{figure}

\begin{figure}
  \centering
  \includegraphics[width=0.45\textwidth]{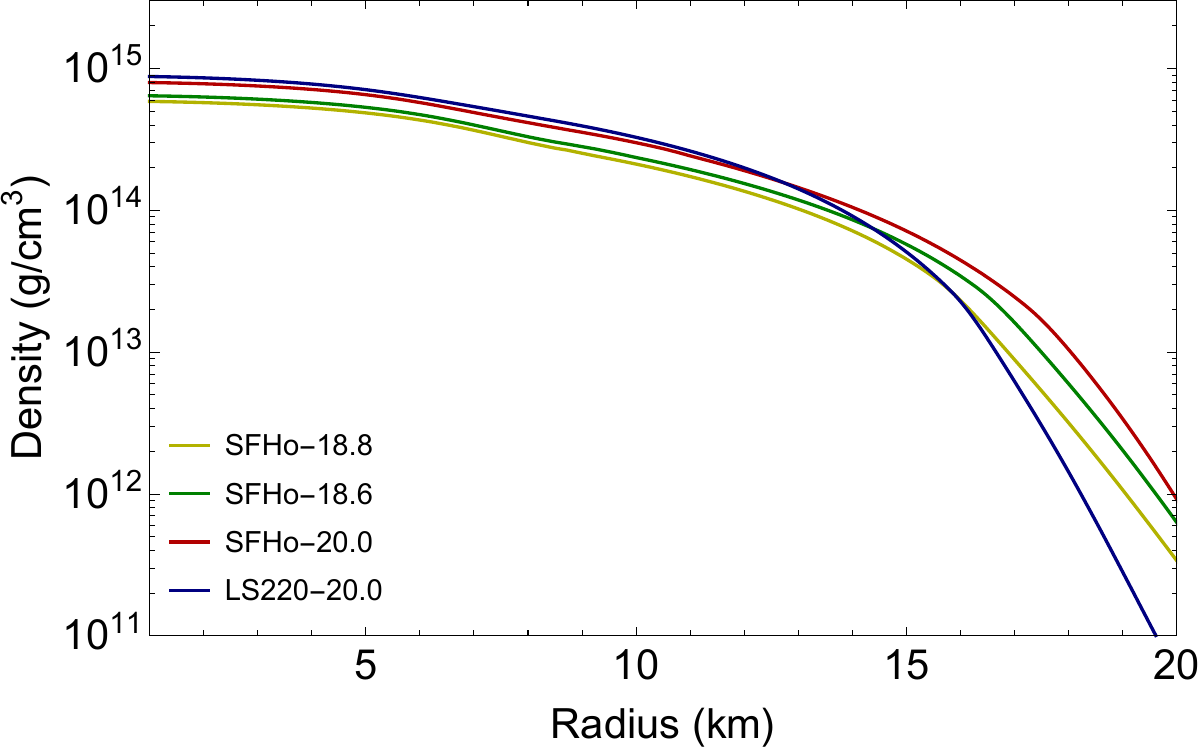}
  \caption{Density profile for the models corresponding to Fig.~\ref{Fig: temp}.
     \label{Fig: density}}
\end{figure}

\begin{figure}
  \centering
  \includegraphics[width=0.45\textwidth]{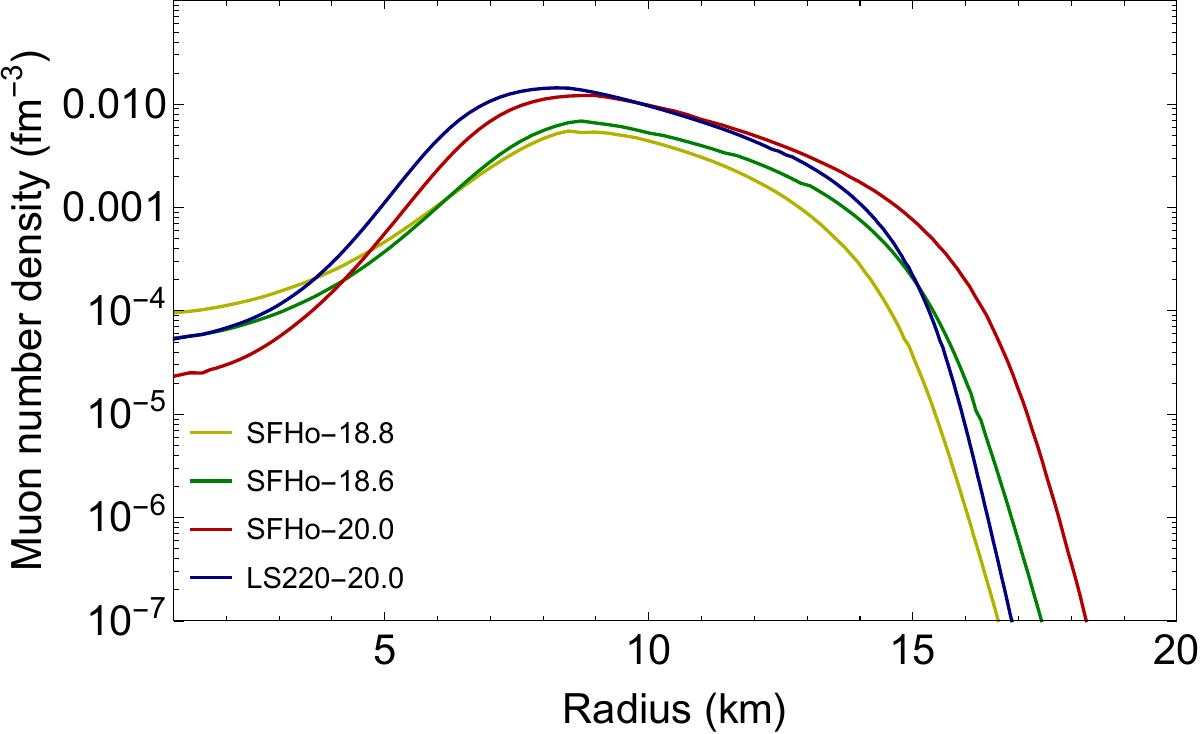}
  \caption{Muon number density. Note that despite large differences in peak temperature, the muon number density does not change by more than an order of magnitude in the relevant region of high muon density (5\,km $\lesssim r \lesssim$ 15\,km).
     \label{Fig: nmu}}
\end{figure}

We have plotted the temperature, density, and muon number density for these simulations at 1 second post-bounce in Figures~\ref{Fig: temp},~\ref{Fig: density}, and~\ref{Fig: nmu} respectively.\footnote{The full profile data can be found at the Garching Core-Collapse Supernova Archive, \url{https://wwwmpa.mpa-garching.mpg.de/ccsnarchive/archive.html}.)}  Note that though the temperature varies by $30\%$, the ultimate muon density, the quantity to which our bound is most sensitive, does not change by more than an $\mathcal{O}(1)$ factor in regions of interest, demonstrating a considerable robustness to large changes in initial parameters. In order to place conservative limits, we ultimately adopt the SFHo-18.8 result as our fiducial profile, which is the coolest and results in the weakest constraint. Additionally, though we place our bound at $t = 1$ sec, we have also computed the bound at $t = 0.5$ sec and $t = 3$ sec, which span the time interval of highest temperature and neutrino luminosity, and find that the ultimate limit on $g_{a\mu}$ does not change by more than a factor of two.

\section{Axion production by muons}
\label{sec:calc}

There are two dominant contributions to the axion emissivity due to muons in the supernova: Compton scattering ($\gamma + \mu \rightarrow a + \mu$) and muon-proton bremsstrahlung ($\mu + p \rightarrow \mu + p + a$). Contributions from muon-muon bremsstrahlung are subdominant as the muon number density is over an order of magnitude below the proton density, muon-electron bremsstrahlung is suppressed by the muon-electron mass ratio, and other channels such as Primakoff or nuclear bremsstrahlung require additional couplings of the axion. 

\begin{figure}
  \centering
  \includegraphics[width=0.45\textwidth]{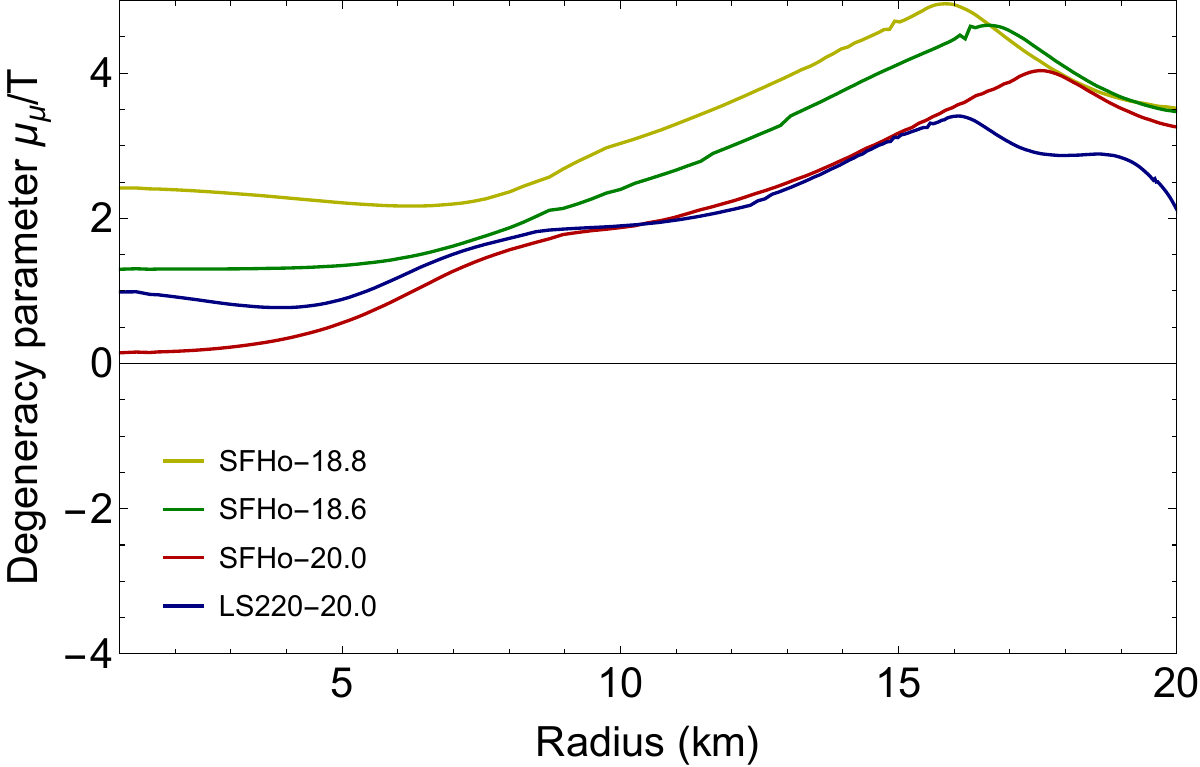}
  \caption{Muon degeneracy parameter $\mu_{\mu}/T$ for various profiles. This ratio never exceeds a small $\mathcal{O}(1)$ value in the relevant region of high muon density (5\,km $\lesssim r \lesssim$ 15\,km).
     \label{Fig: degen}}
\end{figure}

While electrons in the core of the PNS are highly degenerate, suppressing the Compton process~\cite{PhysRevD.33.897} and resulting in bremsstrahlung~\cite{Raffelt:1996wa} as the dominant axion-production channel, muons are only mildly degenerate. This reduces the suppression considerably, allowing the Compton process to become the dominant contribution to axion production. This mild degeneracy is displayed in Figure~\ref{Fig: degen}, where we plot the ratio $\mu_{\mu}/T$ as a function of radius for the various profiles considered. Axion production is only effective in regions of high muon density (5\,km $\lesssim r \lesssim$ 15\,km), where we observe that the degeneracy parameter remains a small $\mathcal{O}(1)$ value.

In order to quantify the effects of the degeneracy on the Compton rate, we compute the approximate suppression factor $F_{\text{deg}}$ by averaging over the Pauli blocking factor (see Sec. 3.2.5 of Ref.~\cite{Raffelt:1996wa}):
\be
F_{\text{deg}} \equiv \frac{1}{n_{\mu}^{\text{th}}} \int\frac{2~d^3\mathbf{p}}{(2\pi)^3}\, f_\mu(E)\left(1- f_\mu(E)\right)\,,
\ee
where $n_{\mu}^{\text{th}}$ is the thermal number density and $f_\mu(E) = (e^{(E-\mu_{\mu})/T}+1)^{-1}$ the energy distribution of muons at the given degeneracy and temperature. Due to the mild degeneracy, this suppression is never in excess of $\sim 15\%$ (see Supplemental Material for a plot of $F_{\text{deg}}$ as a function of radius).

We also compute the degeneracy parameter for the protons, as previous work has suggested that proton degeneracy can play a significant role in the core~\cite{Payez:2014xsa}, but find that protons are solely degenerate for $r \lesssim 4$ km, which is outside the region of high muon abundance, hence proton degeneracy effects are neglected.

Given the above results and the non-relativistic velocities of the muons, the production rate for the Compton process is~\cite{Redondo:2013wwa}\footnote{\label{fn:compton}A word of caution: in a previous version of this paper, the following equation was missing a factor of two. This can be traced to Eq. (2.19) in Ref.~\cite{Redondo:2013wwa}, which does not include the sum over two photon polarizations. Ref.~\cite{Redondo:2013wwa} explicitly states below Eq. (2.9) that these rates must be multiplied by the corresponding factor of two in order to yield the total production rate, however this step has been overlooked in much associated literature. Additionally, in a previous version, the term in brackets was taken in the limit $\omega \ll m_{\mu}$, which is not appropriate here as the typical photon energy is $\omega \sim 3T \sim m_{\mu}$. The inclusion of this factor weakened the initially presented bounds by a factor of 5. A note addressing this appears as an Erratum in the published work.}
\begin{multline}
\label{eq:compton}
\Gamma^{\text{Compton}} = \frac{2\alpha (2 g_{a\mu}m_{\mu})^2}{4 m_{\mu}^2} \\
 \times \left[\frac{\log(1+2x)}{2x} - \frac{1+3x}{(1+2x)^2}\right]
 \frac{n_{\mu}}{e^{\omega/T} - 1} F_{\text{deg}}\,
\end{multline}
with $x \equiv \omega/m_{\mu}$  and $\omega$ the energy of the emitted axion. For muon-proton bremsstrahlung, we have~\cite{Redondo:2013wwa}
\be
\Gamma^{\text{brem}} = \alpha^2 (2 g_{a\mu}m_{\mu})^2 \frac{8\pi}{3\sqrt{2\pi}}\frac{n_p n_{\mu}}{\sqrt{T} m_{\mu}^{7/2} \omega} e^{-\omega/T} F(w, y)
\ee
with
\be
F(w, y) = \int_0^{\infty}~d\tilde{x}\,\tilde{x} e^{-\tilde{x}^2} \int_{\sqrt{\tilde{x}^2 + w} -\tilde{x}}^{\sqrt{\tilde{x}^2 + w}+\tilde{x}}d\tilde{t} \frac{\tilde{t}^3}{(\tilde{t}^2 + y^2)^2} \,,
\ee
where we have defined $w \equiv \omega/T$ and $y \equiv k_S/\sqrt{2 m_{\mu} T}$ with $k_S$ the Debye screening scale, which is the appropriate scale to use for non-degenerate conditions, and $\tilde{x}, \tilde{t}$ are integration variables.

There are two relevant regimes of the $g_{a\mu}$ parameter space: the \textit{free-streaming regime} (small $g_{a\mu}$) and the \textit{trapping regime} (large $g_{a\mu}$). In the free-streaming regime, the mean-free path for axion absorption is significantly larger than the scale of the PNS, hence axions can escape the inner regions of the star without any further interactions, leading to volume emission. At sufficiently strong couplings, however, we enter the trapping regime. The mean free path becomes small in the interior of the PNS and axions are rapidly produced and reabsorbed out to some radius where the mean free path grows long and they escape. This radius is known as the ``axion sphere''~\cite{PhysRevD.42.3297} and the luminosity can be approximated by blackbody emission from that radius at the local temperature~\cite{Raffelt:1996wa}.

Let us begin with the free-streaming regime. In order to compute the free-streaming luminosity of axions due to the muon coupling, we use the following~\cite{Chang:2018rso}:
\begin{multline}
\label{eq:lumi}
L_a^{\text{free}} = \int~dV \int \frac{(4\pi\omega^2)~d\omega}{(2\pi)^3}\\e^{-\tau(\omega,r)} \omega (\Gamma^{\text{Compton}}(\omega, r) + \Gamma^{\text{brem}}(\omega, r))\,,
\end{multline}
with $\tau$ the optical depth\footnote{The appropriate optical depth in this expression is, in contrast to what appeared in an earlier version of this work, calculated using the \textit{reduced} absorptive widths (see Footnote~\ref{fn:opac}). Additionally, the exponential factor in Eq.~\ref{eq:lumi} should be averaged over trajectories, however, in the free-streaming regime, this factor is very near one and hence the differences are negligible.}, given by 
\begin{multline}
\tau(\omega, r) = \int_r^{\infty} dr'~(\Gamma^{\text{Compton}}(\omega, r')+\Gamma^{\text{brem}}(\omega, r'))\\
\times (e^{\omega/T})(1-e^{-\omega/T}) \,,
\end{multline}
where the factor of $e^{\omega/T}$ appears since we are now considering the absorptive widths of the processes, which are related to the production rates by detailed balance: $\Gamma_{\text{prod}} = e^{-\omega/T}\Gamma_{\text{abs}}$, and the factor of $(1-e^{-\omega/T})$ appears as a result of using the reduced absorption rate (see Footnote~\ref{fn:opac}). We include the factor of $e^{-\tau(\omega,r)}$ in the computation of the luminosity to account for the moderate reabsorption of axions by muons even in the ``free-streaming'' regime.\footnote{A very early version of this paper contained an additional factor of $2\omega$ in the denominator of Eq.~\ref{eq:lumi}, which was in error. This caused a factor of one hundred suppression in our bounds. We thank the authors of Ref.~\cite{sam_muon} for bringing this to our attention.}

In the trapped regime, we must first identify the axion sphere, which is done by computing the reduced Rosseland mean opacity\footnote{\label{fn:opac}There are two different procedures being performed in the computation of the following equation. The first is the usual Rosseland mean of the opacity. The second is an explicit reduction of the absorption rate due to stimulated emission, which is necessary to compute the opacity in the context of radiative transfer. The reduction procedure introduces an additional factor of $e^w / (e^w -1)$ in the integrand, as the \textit{reduced absorption rate} is given by $\Gamma^{\text{Compton}}_{\text{abs}}\times(1-e^{-w})$. Note as well that in a previous version of this paper, the equation below contained an erroneous additional factor of $1/2$. However, due to a compensating factor of two in Eq.~\ref{eq:compton} (see Footnote~\ref{fn:compton}), the result was unchanged.} for axions, given by
\be
\label{eq:opacity}
\frac{1}{\rho\kappa_a} = \frac{15}{4\pi^4}\int_0^{\infty}~dw~\left(\Gamma^{\text{Compton}}_{\text{abs}}+\Gamma^{\text{brem}}_{\text{abs}}\right)^ {-1} \frac{w^4 e^{2w}}{(e^w -1)^3} \,,
\ee
where $w \equiv \omega/T$. We can then integrate this radially to find the optical depth $\tau_R$:
\be
\label{eq:tau}
\tau_R(r) = \int_r^{\infty}\rho\kappa_a~dr' \,.
\ee
Whenever $\tau_R \gg 1$, the axions are efficiently trapped~\cite{Carenza_2019}. The axion sphere is defined as the radius $r_a$ at which $\tau_R(r_a) = \frac{2}{3}$. The resulting luminosity is then just approximated as blackbody emission from that radius:
\be
L_a^{\text{trapped}} = \frac{\pi^2}{120}(4\pi r_a^2) T^4 \,,
\ee
As this approximation implicitly assumes that all axions decouple at the same radius, it is only applicable in regions where the coupling is strong enough that absorptive processes shut off over a very narrow region. Otherwise, the energy dependence of the absorptive width stretches this single radius of decoupling into an extended decoupling region. At large values of the coupling, the approximation of an axion sphere is applicable. At the lower limit, as we will show below, the trapped regime transitions smoothly into the free-streaming regime while still producing luminosities in excess of the cooling constraint, hence the effect of an extended decoupling region does not affect our overall bound.

\begin{figure}
  \centering
  \includegraphics[width=0.45\textwidth]{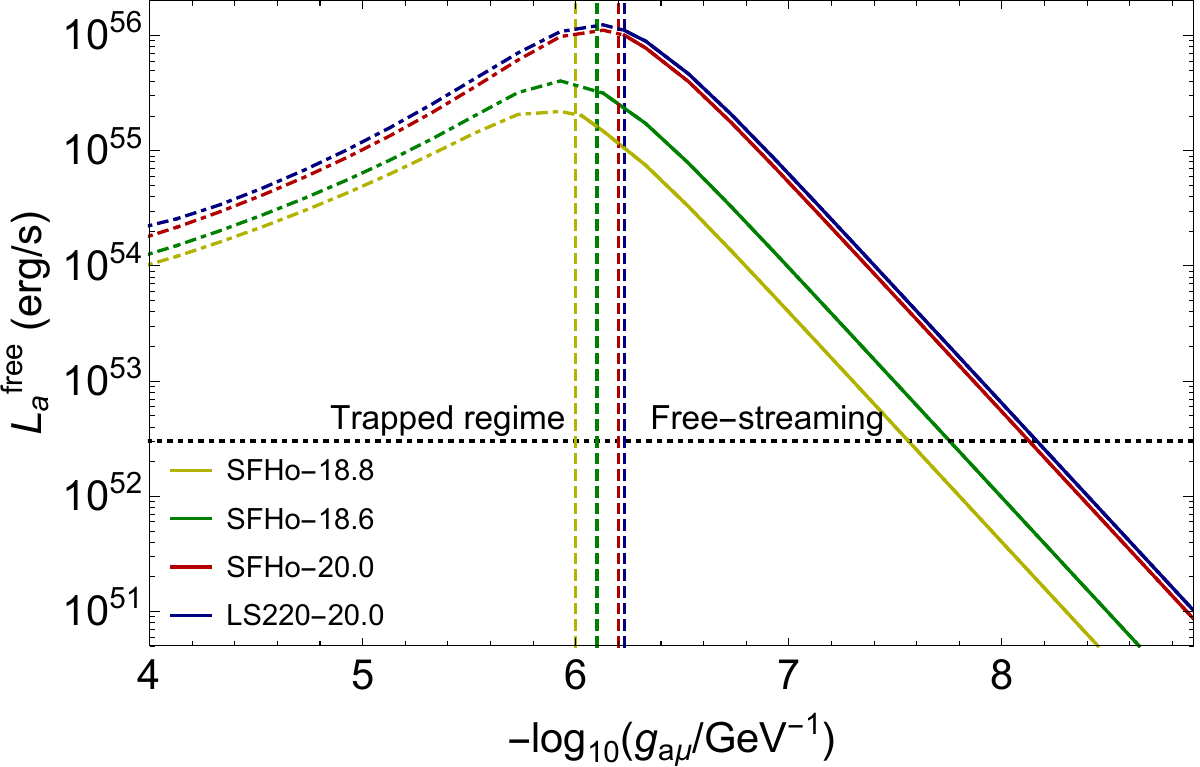}
  \caption{The free-streaming luminosity in axions as a function of axion-muon coupling. Couplings for which the curves lie above the dotted black line and to the right of the dashed lines are excluded by the free-streaming luminosity, while couplings to the left of the vertical dashed lines are excluded by the blackbody approximation in the trapped regime. The decrease of the free-streaming luminosity towards weak couplings is due to rapidly declining axion production, whereas the decrease towards strong couplings is due to an increasing absorptive width and corresponding suppression by the $e^{-\tau(\omega,r)}$ term. Note that the free-streaming luminosity is not the total luminosity, as it neglects the contribution from blackbody surface emission when the optical depth is large. It is shown to the left of the vertical lines purely to demonstrate the expected fall-off in the free-streaming (volume) emission and has hence been plotted as dot-dashed in that region.
     \label{Fig: lumis}}
\end{figure}

\begin{figure}
  \centering
  \includegraphics[width=0.45\textwidth]{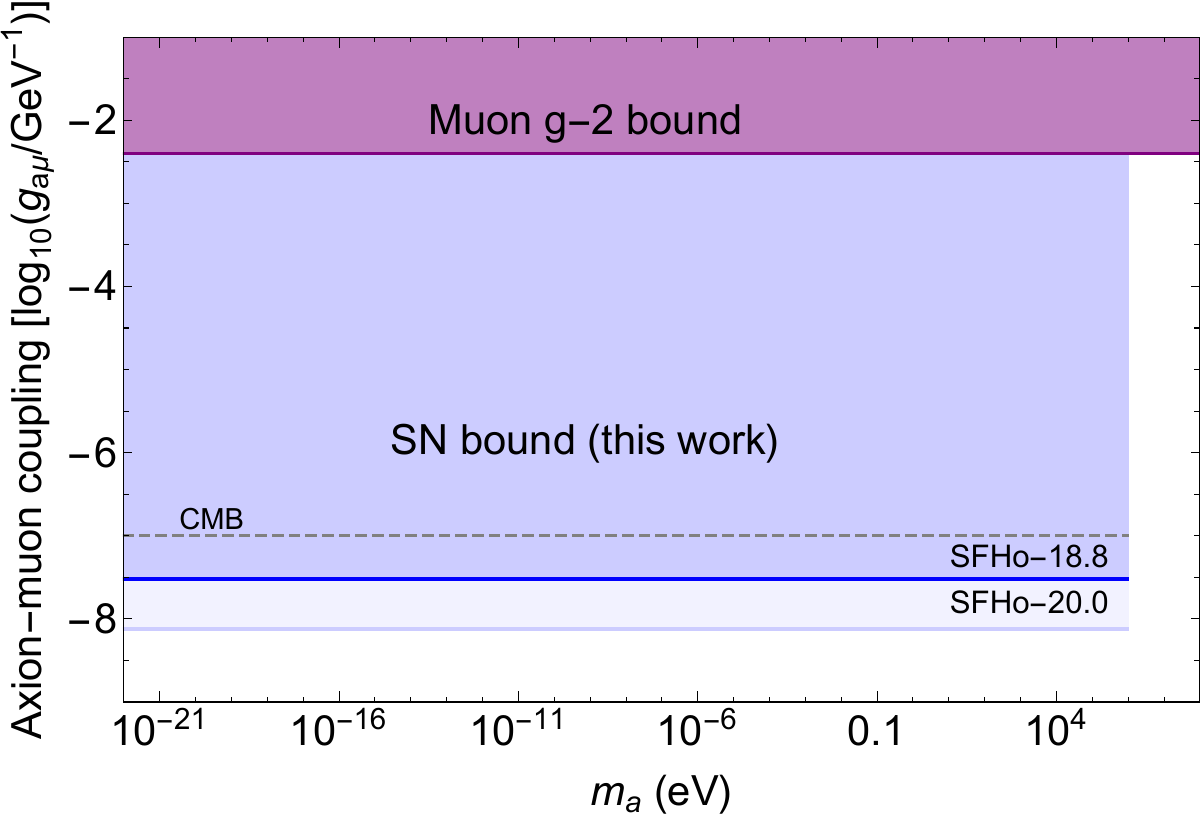}
  \caption{Our constraints, plotted on axion parameter space. Excluded regions are colored. We show the bound that is placed with the SFHo-18.8 profile as a solid blue line, which we quote as our ultimate bound. Note that this is extremely conservative, as model SFHo-18.8 is on the low edge of those compatible with SN 1987A. To illustrate the model dependence, we also show the result for SFHo-20.0 (light blue), which sits on the high edge of the allowed range. We additionally plot existing bounds from virtual axion contributions to the muon $g-2$ (purple), which are the most robust experimental constraints~\cite{Andreas:2010ms}, and a rough estimate of CMB constraints (dashed line, see Supplemental Material for a thorough discussion).
     \label{Fig: bounds}}
\end{figure}

\section{Results}

Figure~\ref{Fig: lumis} shows the results of our luminosity calculations. The curves correspond to the  free-streaming luminosity given in Eq.~\ref{eq:lumi} as a function of coupling, with the horizontal axis showing $-\log_{10} (g_{a\mu}/\text{GeV}^{-1})$. Note that towards weak couplings, the luminosity falls off as the coupling becomes too weak to allow the efficient production of axions. Towards stronger couplings, the free-streaming luminosity also falls off, now due to the increasing absorptive width (i.e. the factor of $e^{-\tau(\omega,r)}$ causes a large suppression).

The vertical lines demarcate our free-streaming regime from our trapped regime. We separate these two regimes at the coupling for which the optical depth within the axion sphere rises above $2/3$. Note that the transition occurs at couplings where we are still excluded by free-streaming, allowing a smooth transition between regimes and no breaks in our bounds. If, instead, the trapped regime had only begun at much stronger couplings, there could have in principle been a gap in the constraint between the two regimes. But this is far from the considered case, hence the actual details of the transition between regimes do not affect our overall results. Note that even at high couplings, the axion sphere is never further out than $\approx 18$ km, as should be evident from the plot of the muon number density (Fig.~\ref{Fig: nmu}). An axion sphere at 18 km corresponds to a blackbody luminosity of $\gtrsim 10^{53}$ erg/s, so our bounds extend smoothly to high couplings.\footnote{At very high couplings, the axion sphere will be pushed to a radius at which the axion luminosity decreases beneath that of the neutrinos. However, the precise location of this transition is subject to considerable uncertainty as it is highly sensitive to the exponential fall-off of the muon abundance. A naive estimate comparing the blackbody luminosity in axions to the canonical $L_{\nu} = 3\times10^{52}$ erg/s places this bound just above the $g-2$ limit, however relying on the one-dimensional simulation results presented here is likely inappropriate, as 3D effects may significantly change the decoupling region. Future work is required to better elucidate the behavior in the very high coupling regime.}

In summary, we can exclude all couplings to the left of the vertical lines (trapped regime) and can exclude all couplings to the right of the vertical line for which the free-streaming luminosity lies above the horizontal line corresponding to $L_a = 3\times10^{52}$ erg/s (free-streaming regime).

These results can be reformulated into a constraint on the axion coupling, the result of which is displayed in Figure~\ref{Fig: bounds}. Translated into a numerical bound, the constraint from SFHo-18.8 is $g_{a\mu} < 10^{-7.5}~\text{GeV}^{-1}$. Note that this is a conservative constraint, as it is placed with the coolest profile, a profile that is on the low-mass edge of the allowed range for SN1987A's remnant. As such, we also show the bound from SFHo-20.0 (lighter blue), which sits on the high-mass edge of the allowed range, though we only quote the conservative bound as our final result.

Additionally, we have cut off our bound at $m_a = 1$\,MeV as our production calculations implicitly took the axion to be effectively massless, a good approximation when $m_a \ll T$, as is the case here. More parameter space (up to $m_a \sim 100$ MeV) could be excluded with a full treatment of a massive axion, though the corresponding Boltzmann suppression will cause the bounds to rise rapidly towards higher masses (see Ref.~\cite{sam_muon}).

In conclusion, SN cooling constrains the axion-muon coupling to
\be
g_{a\mu} < 10^{-7.5}~\text{GeV}^{-1}
\ee
for axions with masses less than an MeV, severely limiting the parameter space for exotic axion-like particles with muon-dominated interactions.

While this analysis was performed in the context of one specific model (an axion with muon-dominated interactions), we wish to note that the general insights presented here open up supernovae as a promising new avenue for robustly probing muonic BSM physics. (For one such extension of this work, see e.g.~\cite{sam_muon}.)

\section*{Acknowledgements}

The authors would like to express our deep appreciation to Djuna Croon, Gilly Elor, Rebecca Leane, and Sam McDermott for alerting us to the mistake in our Eq.~\ref{eq:lumi} and the subsequently-calculated limits. We thank them for reaching out prior to the publication of their paper~\cite{sam_muon}, which significantly extends our work to other couplings and models. Additionally, we would like to thank Ben Wallisch and to thank Dan Green for very useful discussions on the CMB bounds. We are also very grateful to Georg Raffelt for providing valuable insight into subtle errors and inconsistencies in the presentation.

WD and PWG would like to express their gratitude for the support provided by DOE Grant DE-SC0012012, by NSF Grant PHY-1720397, the Heising-Simons Foundation Grants 2015-037 and 2018-0765, DOE HEP QuantISED award \#100495, and the Gordon and Betty Moore Foundation Grant GBMF7946. RB and HTJ acknowledge funding by the European Research Council through Grant ERC-AdG No.~341157-COCO2CASA and by the Deutsche Forschungsgemeinschaft (DFG, German Research Foundation) through Sonderforschungsbereich (Collaborative Research Centre) SFB-1258 ``Neutrinos and Dark Matter in Astro- and Particle Physics (NDM)'' and under Germany's Excellence Strategy through Cluster of Excellence ORIGINS (EXC-2094)---390783311.

\begin{appendix}

\section{Supplemental Material}

\subsection{Comparison to bounds from $N_{\text{eff}}$}
\label{sec:cmb}

Limits on the axion-muon coupling from CMB measurements have also appeared in the literature~\cite{Brust:2013xpv,Baumann:2016wac,DEramo:2018vss}. The physics behind these limits is that muons may produce axions in the early universe, increasing the effective number of relativistic species at the time of the CMB, pushing $N_{\text{eff}}$ above the measured value.

There is a very rapid turnoff in $\Delta N_{\text{eff}}$ as a function of axion-muon coupling \cite{Baumann:2016wac, DEramo:2018vss}.
This can be understood from the fact that once the coupling becomes sufficiently strong that the axions reach thermal equilibrium with the Standard Model bath, their contribution approaches a fixed value around $\Delta N_{\text{eff}} \approx 0.5$. This rapid turn-off has a critical impact on computing bounds---it means that while CMB observations can place a strong bound on the axion-muon coupling if the uncertainty on the $N_{\text{eff}}$ measurement is $\lesssim 0.5$, the bounds disappear essentially entirely if the uncertainty is $\gtrsim 0.5$.  This introduces considerable uncertainty on the bound on the axion coupling when the bounds on $\Delta N_{\text{eff}}$ are around 0.5.

The current precise bound on the axion-muon coupling found in the CMB literature from running a Boltzmann code is around the level of $g_{a\mu} \lesssim 10^{-4}~\text{GeV}^{-1}$~\cite{DEramo:2018vss}.  However using the results from the same paper (Figure 2 of Ref.~\cite{DEramo:2018vss}) we roughly estimate that a significantly stronger bound could be placed~\cite{Priv_Comm}.  But at the same time we must mention that depending on what is believed about current error bars on $\Delta N_{\text{eff}}$, the bound may be significantly weakened, which may be the reason the authors of \cite{DEramo:2018vss} chose to quote a much weaker bound.  Either way, CMB S4 should clear the situation up by providing much smaller uncertainty on $\Delta N_{\text{eff}}$~\cite{Baumann:2016wac}.  Using Figure 2 of Ref.~\cite{DEramo:2018vss} and the value $\Delta N_{\text{eff}} \approx 0.34$ from Ref.~\cite{Aghanim:2018eyx}, we find $g_{a\mu} \approx 10^{-7}~\gev^{-1}$ for $m_a \ll m_{\mu}$. This should \textit{not} be considered a strict limit, however, as a detailed analysis would be required in order to place a robust bound. We have chosen to plot it as a dashed line in Fig. 7 to indicate this.

Depending on the assumptions used when computing $\Delta N_{\text{eff}}$, the $2\sigma$ uncertainty ranges from $0.34$ (Eq. 67b of~\cite{Aghanim:2018eyx}) under the assumption of a pure $\Lambda$CDM model and BBN calculations for the helium fraction $Y_p$ to $0.58$ when $Y_p$ floats independently of $N_{\text{eff}}$ and stellar measurements are not considered (Eq. 82 of~\cite{Aghanim:2018eyx})~\cite{Priv_Comm}. This means that depending on the assumptions made in the calculation of $N_{\text{eff}}$, the bound on the axion-muon coupling could either be strong or non-existent.
The favored measurement of $\Delta N_{\text{eff}}$, using pure $\Lambda$CDM, possesses an uncertainty at $2\sigma$ of $\sim0.34$ and a central value slightly below zero, which would make $\Delta N_{\text{eff}} = 0.5$ a roughly $\sim3.3\sigma$ deviation from the measured value. However, this deviation could be reduced by any modifications to a cosmological model beyond $\Lambda$CDM, e.g. to explain the ongoing Hubble tension, which shift the measurement of $N_{\text{eff}}$ upwards.  As noted, this would have a significant effect on the bounds.

Additionally, it should be noted that independent of the above considerations, if the reheating temperature of the universe was below $m_{\mu}$, these bounds would be greatly weakened.  Of course this is a small part of the currently allowed range for the reheating temperature.  But if, for example, a confirmed positive signal was observed in laboratory experiments and CMB analysis indicated small $\Delta N_{\text{eff}}$, then we would likely not declare a contradiction but instead take this as interesting evidence for a low reheating scale.  In that sense, this CMB bound has some model-dependence.

Taken as a whole, our SN bounds and our estimates of the CMB bounds appear roughly comparable. They are complementary in the sense that the uncertainties and model dependencies in each analysis are completely orthogonal to one another, hence this region of parameter space is strongly disfavored by entirely independent physics.

\begin{figure}[H]
  \centering
  \includegraphics[width=0.45\textwidth]{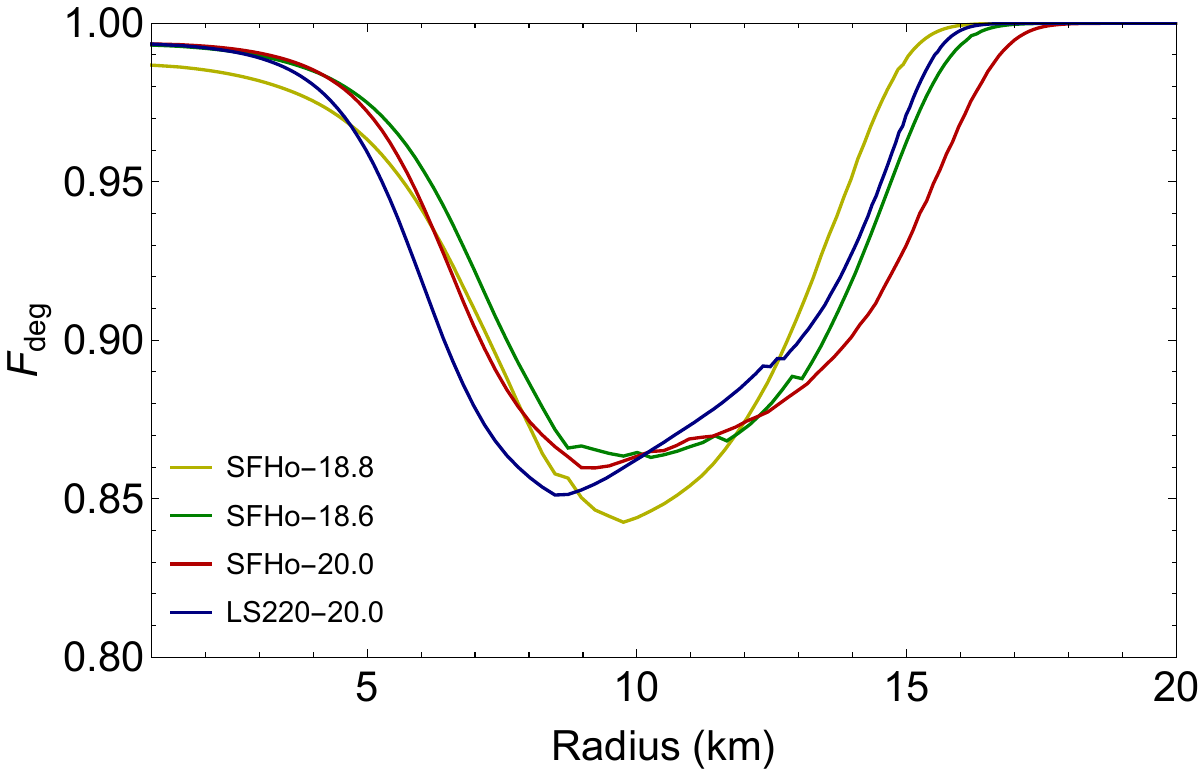}
  \caption{Suppression due to muon degeneracy effects.
       \label{Fig: fdeg}}
\end{figure}

\subsection{Plot of $F_{\text{deg}}$}

As is explained in the text, we quantify the effects of muon degeneracy on the Compton rate by computing the suppression factor $F_{\text{deg}}$ (see Sec. 3.2.5 of Ref.~\cite{Raffelt:1996wa}):
\be
F_{\text{deg}} \equiv \frac{1}{n_{\mu}^{\text{th}}} \int\frac{2~d^3\mathbf{p}}{(2\pi)^3}\, f_\mu(E)\left(1- f_\mu(E)\right)\,,
\ee
where $n_{\mu}^{\text{th}}$ is the thermal number density and $f_\mu(E) = (e^{(E-\mu_{\mu})/T}+1)^{-1}$ the energy distribution of muons at the given degeneracy and temperature. We have plotted this suppression factor in Figure~\ref{Fig: fdeg}. Due to the mild degeneracy, the suppression is never in excess of $\sim 15\%$.

\end{appendix}

\bibliography{ref}

\begin{thebibliography}{61}
\expandafter\ifx\csname natexlab\endcsname\relax\def\natexlab#1{#1}\fi
\expandafter\ifx\csname bibnamefont\endcsname\relax
  \def\bibnamefont#1{#1}\fi
\expandafter\ifx\csname bibfnamefont\endcsname\relax
  \def\bibfnamefont#1{#1}\fi
\expandafter\ifx\csname citenamefont\endcsname\relax
  \def\citenamefont#1{#1}\fi
\expandafter\ifx\csname url\endcsname\relax
  \def\url#1{\texttt{#1}}\fi
\expandafter\ifx\csname urlprefix\endcsname\relax\def\urlprefix{URL }\fi
\providecommand{\bibinfo}[2]{#2}
\providecommand{\eprint}[2][]{\url{#2}}

\bibitem[{\citenamefont{Peccei and Quinn}(1977{\natexlab{a}})}]{Peccei:1977hh}
\bibinfo{author}{\bibfnamefont{R.~D.} \bibnamefont{Peccei}} \bibnamefont{and}
  \bibinfo{author}{\bibfnamefont{H.~R.} \bibnamefont{Quinn}},
  \bibinfo{journal}{Phys. Rev. Lett.} \textbf{\bibinfo{volume}{38}},
  \bibinfo{pages}{1440} (\bibinfo{year}{1977}{\natexlab{a}}),
  \bibinfo{note}{[,328(1977)]}.

\bibitem[{\citenamefont{Peccei and Quinn}(1977{\natexlab{b}})}]{Peccei:1977ur}
\bibinfo{author}{\bibfnamefont{R.~D.} \bibnamefont{Peccei}} \bibnamefont{and}
  \bibinfo{author}{\bibfnamefont{H.~R.} \bibnamefont{Quinn}},
  \bibinfo{journal}{Phys. Rev.} \textbf{\bibinfo{volume}{D16}},
  \bibinfo{pages}{1791} (\bibinfo{year}{1977}{\natexlab{b}}).

\bibitem[{\citenamefont{Weinberg}(1978)}]{Weinberg:1977ma}
\bibinfo{author}{\bibfnamefont{S.}~\bibnamefont{Weinberg}},
  \bibinfo{journal}{Phys. Rev. Lett.} \textbf{\bibinfo{volume}{40}},
  \bibinfo{pages}{223} (\bibinfo{year}{1978}).

\bibitem[{\citenamefont{Arvanitaki et~al.}(2010)\citenamefont{Arvanitaki,
  Dimopoulos, Dubovsky, Kaloper, and March-Russell}}]{Arvanitaki:2009fg}
\bibinfo{author}{\bibfnamefont{A.}~\bibnamefont{Arvanitaki}},
  \bibinfo{author}{\bibfnamefont{S.}~\bibnamefont{Dimopoulos}},
  \bibinfo{author}{\bibfnamefont{S.}~\bibnamefont{Dubovsky}},
  \bibinfo{author}{\bibfnamefont{N.}~\bibnamefont{Kaloper}}, \bibnamefont{and}
  \bibinfo{author}{\bibfnamefont{J.}~\bibnamefont{March-Russell}},
  \bibinfo{journal}{Phys. Rev.} \textbf{\bibinfo{volume}{D81}},
  \bibinfo{pages}{123530} (\bibinfo{year}{2010}), \eprint{0905.4720}.

\bibitem[{\citenamefont{Preskill et~al.}(1983)\citenamefont{Preskill, Wise, and
  Wilczek}}]{Preskill:1982cy}
\bibinfo{author}{\bibfnamefont{J.}~\bibnamefont{Preskill}},
  \bibinfo{author}{\bibfnamefont{M.~B.} \bibnamefont{Wise}}, \bibnamefont{and}
  \bibinfo{author}{\bibfnamefont{F.}~\bibnamefont{Wilczek}},
  \bibinfo{journal}{Phys. Lett. B} \textbf{\bibinfo{volume}{120}},
  \bibinfo{pages}{127} (\bibinfo{year}{1983}).

\bibitem[{\citenamefont{Abbott and Sikivie}(1983)}]{Abbott:1982af}
\bibinfo{author}{\bibfnamefont{L.}~\bibnamefont{Abbott}} \bibnamefont{and}
  \bibinfo{author}{\bibfnamefont{P.}~\bibnamefont{Sikivie}},
  \bibinfo{journal}{Phys. Lett. B} \textbf{\bibinfo{volume}{120}},
  \bibinfo{pages}{133} (\bibinfo{year}{1983}).

\bibitem[{\citenamefont{Dine and Fischler}(1983)}]{Dine:1982ah}
\bibinfo{author}{\bibfnamefont{M.}~\bibnamefont{Dine}} \bibnamefont{and}
  \bibinfo{author}{\bibfnamefont{W.}~\bibnamefont{Fischler}},
  \bibinfo{journal}{Phys. Lett. B} \textbf{\bibinfo{volume}{120}},
  \bibinfo{pages}{137} (\bibinfo{year}{1983}).

\bibitem[{\citenamefont{Duffy and van Bibber}(2009)}]{Duffy:2009ig}
\bibinfo{author}{\bibfnamefont{L.~D.} \bibnamefont{Duffy}} \bibnamefont{and}
  \bibinfo{author}{\bibfnamefont{K.}~\bibnamefont{van Bibber}},
  \bibinfo{journal}{New J. Phys.} \textbf{\bibinfo{volume}{11}},
  \bibinfo{pages}{105008} (\bibinfo{year}{2009}), \eprint{0904.3346}.

\bibitem[{\citenamefont{Ringwald}(2016)}]{Ringwald:2016yge}
\bibinfo{author}{\bibfnamefont{A.}~\bibnamefont{Ringwald}},
  \bibinfo{journal}{PoS} \textbf{\bibinfo{volume}{NOW2016}},
  \bibinfo{pages}{081} (\bibinfo{year}{2016}), \eprint{1612.08933}.

\bibitem[{\citenamefont{Anastassopoulos
  et~al.}(2017)}]{Anastassopoulos:2017ftl}
\bibinfo{author}{\bibfnamefont{V.}~\bibnamefont{Anastassopoulos}}
  \bibnamefont{et~al.} (\bibinfo{collaboration}{CAST}),
  \bibinfo{journal}{Nature Phys.} \textbf{\bibinfo{volume}{13}},
  \bibinfo{pages}{584} (\bibinfo{year}{2017}), \eprint{1705.02290}.

\bibitem[{\citenamefont{Du et~al.}(2018)\citenamefont{Du, Force, Khatiwada,
  Lentz, Ottens, Rosenberg, Rybka, Carosi, Woollett, Bowring
  et~al.}}]{PhysRevLett.120.151301}
\bibinfo{author}{\bibfnamefont{N.}~\bibnamefont{Du}},
  \bibinfo{author}{\bibfnamefont{N.}~\bibnamefont{Force}},
  \bibinfo{author}{\bibfnamefont{R.}~\bibnamefont{Khatiwada}},
  \bibinfo{author}{\bibfnamefont{E.}~\bibnamefont{Lentz}},
  \bibinfo{author}{\bibfnamefont{R.}~\bibnamefont{Ottens}},
  \bibinfo{author}{\bibfnamefont{L.~J.} \bibnamefont{Rosenberg}},
  \bibinfo{author}{\bibfnamefont{G.}~\bibnamefont{Rybka}},
  \bibinfo{author}{\bibfnamefont{G.}~\bibnamefont{Carosi}},
  \bibinfo{author}{\bibfnamefont{N.}~\bibnamefont{Woollett}},
  \bibinfo{author}{\bibfnamefont{D.}~\bibnamefont{Bowring}},
  \bibnamefont{et~al.} (\bibinfo{collaboration}{ADMX Collaboration}),
  \bibinfo{journal}{Phys. Rev. Lett.} \textbf{\bibinfo{volume}{120}},
  \bibinfo{pages}{151301} (\bibinfo{year}{2018}),
  \urlprefix\url{https://link.aps.org/doi/10.1103/PhysRevLett.120.151301}.

\bibitem[{\citenamefont{Brubaker et~al.}(2017)}]{Brubaker:2016ktl}
\bibinfo{author}{\bibfnamefont{B.~M.} \bibnamefont{Brubaker}}
  \bibnamefont{et~al.}, \bibinfo{journal}{Phys. Rev. Lett.}
  \textbf{\bibinfo{volume}{118}}, \bibinfo{pages}{061302}
  (\bibinfo{year}{2017}), \eprint{1610.02580}.

\bibitem[{\citenamefont{Betz et~al.}(2013)\citenamefont{Betz, Caspers, Gasior,
  Thumm, and Rieger}}]{Betz:2013dza}
\bibinfo{author}{\bibfnamefont{M.}~\bibnamefont{Betz}},
  \bibinfo{author}{\bibfnamefont{F.}~\bibnamefont{Caspers}},
  \bibinfo{author}{\bibfnamefont{M.}~\bibnamefont{Gasior}},
  \bibinfo{author}{\bibfnamefont{M.}~\bibnamefont{Thumm}}, \bibnamefont{and}
  \bibinfo{author}{\bibfnamefont{S.~W.} \bibnamefont{Rieger}},
  \bibinfo{journal}{Phys. Rev.} \textbf{\bibinfo{volume}{D88}},
  \bibinfo{pages}{075014} (\bibinfo{year}{2013}), \eprint{1310.8098}.

\bibitem[{\citenamefont{Barth et~al.}(2013)\citenamefont{Barth, Belov, Beltran,
  Bräuninger, Carmona, Collar, Dafni, Davenport, Lella, Eleftheriadis
  et~al.}}]{Barth_2013}
\bibinfo{author}{\bibfnamefont{K.}~\bibnamefont{Barth}},
  \bibinfo{author}{\bibfnamefont{A.}~\bibnamefont{Belov}},
  \bibinfo{author}{\bibfnamefont{B.}~\bibnamefont{Beltran}},
  \bibinfo{author}{\bibfnamefont{H.}~\bibnamefont{Bräuninger}},
  \bibinfo{author}{\bibfnamefont{J.}~\bibnamefont{Carmona}},
  \bibinfo{author}{\bibfnamefont{J.}~\bibnamefont{Collar}},
  \bibinfo{author}{\bibfnamefont{T.}~\bibnamefont{Dafni}},
  \bibinfo{author}{\bibfnamefont{M.}~\bibnamefont{Davenport}},
  \bibinfo{author}{\bibfnamefont{L.~D.} \bibnamefont{Lella}},
  \bibinfo{author}{\bibfnamefont{C.}~\bibnamefont{Eleftheriadis}},
  \bibnamefont{et~al.}, \bibinfo{journal}{Journal of Cosmology and
  Astroparticle Physics} \textbf{\bibinfo{volume}{2013}}, \bibinfo{pages}{010}
  (\bibinfo{year}{2013}),
  \urlprefix\url{https://doi.org/10.1088%2F1475-7516%2F2013%2F05%2F010}.

\bibitem[{\citenamefont{Jackson~Kimball et~al.}(2017)}]{JacksonKimball:2017elr}
\bibinfo{author}{\bibfnamefont{D.~F.} \bibnamefont{Jackson~Kimball}}
  \bibnamefont{et~al.} (\bibinfo{year}{2017}), \eprint{1711.08999}.

\bibitem[{\citenamefont{Graham et~al.}(2015)\citenamefont{Graham, Irastorza,
  Lamoreaux, Lindner, and van Bibber}}]{Graham:2015ouw}
\bibinfo{author}{\bibfnamefont{P.~W.} \bibnamefont{Graham}},
  \bibinfo{author}{\bibfnamefont{I.~G.} \bibnamefont{Irastorza}},
  \bibinfo{author}{\bibfnamefont{S.~K.} \bibnamefont{Lamoreaux}},
  \bibinfo{author}{\bibfnamefont{A.}~\bibnamefont{Lindner}}, \bibnamefont{and}
  \bibinfo{author}{\bibfnamefont{K.~A.} \bibnamefont{van Bibber}},
  \bibinfo{journal}{Ann. Rev. Nucl. Part. Sci.} \textbf{\bibinfo{volume}{65}},
  \bibinfo{pages}{485} (\bibinfo{year}{2015}), \eprint{1602.00039}.

\bibitem[{\citenamefont{Lohs et~al.}(2015)\citenamefont{Lohs, Martinez-Pinedo,
  and Fischer}}]{Lohs:2015qyn}
\bibinfo{author}{\bibfnamefont{A.}~\bibnamefont{Lohs}},
  \bibinfo{author}{\bibfnamefont{G.}~\bibnamefont{Martinez-Pinedo}},
  \bibnamefont{and} \bibinfo{author}{\bibfnamefont{T.}~\bibnamefont{Fischer}},
  \bibinfo{journal}{PoS} \textbf{\bibinfo{volume}{NICXIII}},
  \bibinfo{pages}{118} (\bibinfo{year}{2015}).

\bibitem[{\citenamefont{Brust et~al.}(2013)\citenamefont{Brust, Kaplan, and
  Walters}}]{Brust:2013xpv}
\bibinfo{author}{\bibfnamefont{C.}~\bibnamefont{Brust}},
  \bibinfo{author}{\bibfnamefont{D.~E.} \bibnamefont{Kaplan}},
  \bibnamefont{and} \bibinfo{author}{\bibfnamefont{M.~T.}
  \bibnamefont{Walters}}, \bibinfo{journal}{JHEP}
  \textbf{\bibinfo{volume}{12}}, \bibinfo{pages}{058} (\bibinfo{year}{2013}),
  \eprint{1303.5379}.

\bibitem[{\citenamefont{Bollig et~al.}(2017)\citenamefont{Bollig, Janka, Lohs,
  Martinez-Pinedo, Horowitz, and Melson}}]{Bollig:2017lki}
\bibinfo{author}{\bibfnamefont{R.}~\bibnamefont{Bollig}},
  \bibinfo{author}{\bibfnamefont{H.~T.} \bibnamefont{Janka}},
  \bibinfo{author}{\bibfnamefont{A.}~\bibnamefont{Lohs}},
  \bibinfo{author}{\bibfnamefont{G.}~\bibnamefont{Martinez-Pinedo}},
  \bibinfo{author}{\bibfnamefont{C.~J.} \bibnamefont{Horowitz}},
  \bibnamefont{and} \bibinfo{author}{\bibfnamefont{T.}~\bibnamefont{Melson}},
  \bibinfo{journal}{Phys. Rev. Lett.} \textbf{\bibinfo{volume}{119}},
  \bibinfo{pages}{242702} (\bibinfo{year}{2017}), \eprint{1706.04630}.

\bibitem[{\citenamefont{Andreas et~al.}(2010)\citenamefont{Andreas, Lebedev,
  Ramos-Sanchez, and Ringwald}}]{Andreas:2010ms}
\bibinfo{author}{\bibfnamefont{S.}~\bibnamefont{Andreas}},
  \bibinfo{author}{\bibfnamefont{O.}~\bibnamefont{Lebedev}},
  \bibinfo{author}{\bibfnamefont{S.}~\bibnamefont{Ramos-Sanchez}},
  \bibnamefont{and} \bibinfo{author}{\bibfnamefont{A.}~\bibnamefont{Ringwald}},
  \bibinfo{journal}{JHEP} \textbf{\bibinfo{volume}{08}}, \bibinfo{pages}{003}
  (\bibinfo{year}{2010}), \eprint{1005.3978}.

\bibitem[{\citenamefont{Baumann et~al.}(2016)\citenamefont{Baumann, Green, and
  Wallisch}}]{Baumann:2016wac}
\bibinfo{author}{\bibfnamefont{D.}~\bibnamefont{Baumann}},
  \bibinfo{author}{\bibfnamefont{D.}~\bibnamefont{Green}}, \bibnamefont{and}
  \bibinfo{author}{\bibfnamefont{B.}~\bibnamefont{Wallisch}},
  \bibinfo{journal}{Phys. Rev. Lett.} \textbf{\bibinfo{volume}{117}},
  \bibinfo{pages}{171301} (\bibinfo{year}{2016}), \eprint{1604.08614}.

\bibitem[{\citenamefont{D'Eramo et~al.}(2018)\citenamefont{D'Eramo, Ferreira,
  Notari, and Bernal}}]{DEramo:2018vss}
\bibinfo{author}{\bibfnamefont{F.}~\bibnamefont{D'Eramo}},
  \bibinfo{author}{\bibfnamefont{R.~Z.} \bibnamefont{Ferreira}},
  \bibinfo{author}{\bibfnamefont{A.}~\bibnamefont{Notari}}, \bibnamefont{and}
  \bibinfo{author}{\bibfnamefont{J.~L.} \bibnamefont{Bernal}},
  \bibinfo{journal}{JCAP} \textbf{\bibinfo{volume}{1811}}, \bibinfo{pages}{014}
  (\bibinfo{year}{2018}), \eprint{1808.07430}.

\bibitem[{\citenamefont{Raffelt}(1996)}]{Raffelt:1996wa}
\bibinfo{author}{\bibfnamefont{G.~G.} \bibnamefont{Raffelt}},
  \emph{\bibinfo{title}{{Stars as laboratories for fundamental physics}}}
  (\bibinfo{publisher}{The University of Chicago Press, Chicago},
  \bibinfo{year}{1996}), ISBN \bibinfo{isbn}{9780226702728},
  \urlprefix\url{http://wwwth.mpp.mpg.de/members/raffelt/mypapers/199613.pdf}.

\bibitem[{\citenamefont{Turner}(1988)}]{PhysRevLett.60.1797}
\bibinfo{author}{\bibfnamefont{M.~S.} \bibnamefont{Turner}},
  \bibinfo{journal}{Phys. Rev. Lett.} \textbf{\bibinfo{volume}{60}},
  \bibinfo{pages}{1797} (\bibinfo{year}{1988}),
  \urlprefix\url{https://link.aps.org/doi/10.1103/PhysRevLett.60.1797}.

\bibitem[{\citenamefont{Burrows et~al.}(1989)\citenamefont{Burrows, Turner, and
  Brinkmann}}]{PhysRevD.39.1020}
\bibinfo{author}{\bibfnamefont{A.}~\bibnamefont{Burrows}},
  \bibinfo{author}{\bibfnamefont{M.~S.} \bibnamefont{Turner}},
  \bibnamefont{and} \bibinfo{author}{\bibfnamefont{R.~P.}
  \bibnamefont{Brinkmann}}, \bibinfo{journal}{Phys. Rev. D}
  \textbf{\bibinfo{volume}{39}}, \bibinfo{pages}{1020} (\bibinfo{year}{1989}),
  \urlprefix\url{https://link.aps.org/doi/10.1103/PhysRevD.39.1020}.

\bibitem[{\citenamefont{Raffelt}(1990)}]{RAFFELT19901}
\bibinfo{author}{\bibfnamefont{G.~G.} \bibnamefont{Raffelt}},
  \bibinfo{journal}{Physics Reports} \textbf{\bibinfo{volume}{198}},
  \bibinfo{pages}{1 } (\bibinfo{year}{1990}), ISSN \bibinfo{issn}{0370-1573},
  \urlprefix\url{http://www.sciencedirect.com/science/article/pii/0370157390900546}.

\bibitem[{\citenamefont{Chang et~al.}(2018)\citenamefont{Chang, Essig, and
  McDermott}}]{Chang:2018rso}
\bibinfo{author}{\bibfnamefont{J.~H.} \bibnamefont{Chang}},
  \bibinfo{author}{\bibfnamefont{R.}~\bibnamefont{Essig}}, \bibnamefont{and}
  \bibinfo{author}{\bibfnamefont{S.~D.} \bibnamefont{McDermott}},
  \bibinfo{journal}{JHEP} \textbf{\bibinfo{volume}{09}}, \bibinfo{pages}{051}
  (\bibinfo{year}{2018}), \eprint{1803.00993}.

\bibitem[{\citenamefont{DeRocco et~al.}(2019)\citenamefont{DeRocco, Graham,
  Kasen, Marques-Tavares, and Rajendran}}]{DeRocco:2019jti}
\bibinfo{author}{\bibfnamefont{W.}~\bibnamefont{DeRocco}},
  \bibinfo{author}{\bibfnamefont{P.~W.} \bibnamefont{Graham}},
  \bibinfo{author}{\bibfnamefont{D.}~\bibnamefont{Kasen}},
  \bibinfo{author}{\bibfnamefont{G.}~\bibnamefont{Marques-Tavares}},
  \bibnamefont{and}
  \bibinfo{author}{\bibfnamefont{S.}~\bibnamefont{Rajendran}},
  \bibinfo{journal}{Phys. Rev.} \textbf{\bibinfo{volume}{D100}},
  \bibinfo{pages}{075018} (\bibinfo{year}{2019}), \eprint{1905.09284}.

\bibitem[{\citenamefont{Carenza et~al.}(2019)\citenamefont{Carenza, Fischer,
  Giannotti, Guo, Martinez-Pinedo, and Mirizzi}}]{Carenza_2019}
\bibinfo{author}{\bibfnamefont{P.}~\bibnamefont{Carenza}},
  \bibinfo{author}{\bibfnamefont{T.}~\bibnamefont{Fischer}},
  \bibinfo{author}{\bibfnamefont{M.}~\bibnamefont{Giannotti}},
  \bibinfo{author}{\bibfnamefont{G.}~\bibnamefont{Guo}},
  \bibinfo{author}{\bibfnamefont{G.}~\bibnamefont{Martinez-Pinedo}},
  \bibnamefont{and} \bibinfo{author}{\bibfnamefont{A.}~\bibnamefont{Mirizzi}},
  \bibinfo{journal}{Journal of Cosmology and Astroparticle Physics}
  \textbf{\bibinfo{volume}{2019}}, \bibinfo{pages}{016} (\bibinfo{year}{2019}),
  \urlprefix\url{https://doi.org/10.1088%2F1475-7516%2F2019%2F10%2F016}.

\bibitem[{\citenamefont{{Bar} et~al.}(2019)\citenamefont{{Bar}, {Blum}, and
  {D'Amico}}}]{Bar+2019}
\bibinfo{author}{\bibfnamefont{N.}~\bibnamefont{{Bar}}},
  \bibinfo{author}{\bibfnamefont{K.}~\bibnamefont{{Blum}}}, \bibnamefont{and}
  \bibinfo{author}{\bibfnamefont{G.}~\bibnamefont{{D'Amico}}},
  \bibinfo{journal}{arXiv e-prints} \bibinfo{eid}{arXiv:1907.05020}
  (\bibinfo{year}{2019}), \eprint{1907.05020}.

\bibitem[{\citenamefont{{Blum} and {Kushnir}}(2016)}]{BlumKushnir2016}
\bibinfo{author}{\bibfnamefont{K.}~\bibnamefont{{Blum}}} \bibnamefont{and}
  \bibinfo{author}{\bibfnamefont{D.}~\bibnamefont{{Kushnir}}},
  \bibinfo{journal}{Astrophys. J.} \textbf{\bibinfo{volume}{828}},
  \bibinfo{eid}{31} (\bibinfo{year}{2016}), \eprint{1601.03422}.

\bibitem[{\citenamefont{{Kushnir} and {Katz}}(2015)}]{KushnirKatz2015}
\bibinfo{author}{\bibfnamefont{D.}~\bibnamefont{{Kushnir}}} \bibnamefont{and}
  \bibinfo{author}{\bibfnamefont{B.}~\bibnamefont{{Katz}}},
  \bibinfo{journal}{Astrophys. J.} \textbf{\bibinfo{volume}{811}},
  \bibinfo{eid}{97} (\bibinfo{year}{2015}), \eprint{1412.1096}.

\bibitem[{\citenamefont{{Cigan} et~al.}(2019)\citenamefont{{Cigan}, {Matsuura},
  {Gomez}, {Indebetouw}, {Abell{\'a}n}, {Gabler}, {Richards}, {Alp}, {Davis},
  {Janka} et~al.}}]{Cigan+2019}
\bibinfo{author}{\bibfnamefont{P.}~\bibnamefont{{Cigan}}},
  \bibinfo{author}{\bibfnamefont{M.}~\bibnamefont{{Matsuura}}},
  \bibinfo{author}{\bibfnamefont{H.~L.} \bibnamefont{{Gomez}}},
  \bibinfo{author}{\bibfnamefont{R.}~\bibnamefont{{Indebetouw}}},
  \bibinfo{author}{\bibfnamefont{F.}~\bibnamefont{{Abell{\'a}n}}},
  \bibinfo{author}{\bibfnamefont{M.}~\bibnamefont{{Gabler}}},
  \bibinfo{author}{\bibfnamefont{A.}~\bibnamefont{{Richards}}},
  \bibinfo{author}{\bibfnamefont{D.}~\bibnamefont{{Alp}}},
  \bibinfo{author}{\bibfnamefont{T.~A.} \bibnamefont{{Davis}}},
  \bibinfo{author}{\bibfnamefont{H.-T.} \bibnamefont{{Janka}}},
  \bibnamefont{et~al.}, \bibinfo{journal}{Astrophys. J.}
  \textbf{\bibinfo{volume}{886}}, \bibinfo{eid}{51} (\bibinfo{year}{2019}),
  \eprint{1910.02960}.

\bibitem[{\citenamefont{{Page} et~al.}(2020)\citenamefont{{Page}, {Beznogov},
  {Garibay}, {Lattimer}, {Prakash}, and {Janka}}}]{Page+2020}
\bibinfo{author}{\bibfnamefont{D.}~\bibnamefont{{Page}}},
  \bibinfo{author}{\bibfnamefont{M.~V.} \bibnamefont{{Beznogov}}},
  \bibinfo{author}{\bibfnamefont{I.}~\bibnamefont{{Garibay}}},
  \bibinfo{author}{\bibfnamefont{J.~M.} \bibnamefont{{Lattimer}}},
  \bibinfo{author}{\bibfnamefont{M.}~\bibnamefont{{Prakash}}},
  \bibnamefont{and} \bibinfo{author}{\bibfnamefont{H.-T.}
  \bibnamefont{{Janka}}}, \bibinfo{journal}{arXiv e-prints}
  \bibinfo{eid}{arXiv:2004.06078} (\bibinfo{year}{2020}), \eprint{2004.06078}.

\bibitem[{\citenamefont{{Fischer} et~al.}(2014)\citenamefont{{Fischer},
  {Hempel}, {Sagert}, {Suwa}, and {Schaffner-Bielich}}}]{Fischer+2014}
\bibinfo{author}{\bibfnamefont{T.}~\bibnamefont{{Fischer}}},
  \bibinfo{author}{\bibfnamefont{M.}~\bibnamefont{{Hempel}}},
  \bibinfo{author}{\bibfnamefont{I.}~\bibnamefont{{Sagert}}},
  \bibinfo{author}{\bibfnamefont{Y.}~\bibnamefont{{Suwa}}}, \bibnamefont{and}
  \bibinfo{author}{\bibfnamefont{J.}~\bibnamefont{{Schaffner-Bielich}}},
  \bibinfo{journal}{European Physical Journal A} \textbf{\bibinfo{volume}{50}},
  \bibinfo{eid}{46} (\bibinfo{year}{2014}), \eprint{1307.6190}.

\bibitem[{\citenamefont{{Oertel} et~al.}(2017)\citenamefont{{Oertel}, {Hempel},
  {Kl{\"a}hn}, and {Typel}}}]{Oertel+2017}
\bibinfo{author}{\bibfnamefont{M.}~\bibnamefont{{Oertel}}},
  \bibinfo{author}{\bibfnamefont{M.}~\bibnamefont{{Hempel}}},
  \bibinfo{author}{\bibfnamefont{T.}~\bibnamefont{{Kl{\"a}hn}}},
  \bibnamefont{and} \bibinfo{author}{\bibfnamefont{S.}~\bibnamefont{{Typel}}},
  \bibinfo{journal}{Reviews of Modern Physics} \textbf{\bibinfo{volume}{89}},
  \bibinfo{eid}{015007} (\bibinfo{year}{2017}), \eprint{1610.03361}.

\bibitem[{\citenamefont{{Fischer} et~al.}(2017)\citenamefont{{Fischer},
  {Bastian}, {Blaschke}, {Cierniak}, {Hempel}, {Kl{\"a}hn},
  {Mart{\'\i}nez-Pinedo}, {Newton}, {R{\"o}pke}, and {Typel}}}]{Fischer+2017}
\bibinfo{author}{\bibfnamefont{T.}~\bibnamefont{{Fischer}}},
  \bibinfo{author}{\bibfnamefont{N.-U.} \bibnamefont{{Bastian}}},
  \bibinfo{author}{\bibfnamefont{D.}~\bibnamefont{{Blaschke}}},
  \bibinfo{author}{\bibfnamefont{M.}~\bibnamefont{{Cierniak}}},
  \bibinfo{author}{\bibfnamefont{M.}~\bibnamefont{{Hempel}}},
  \bibinfo{author}{\bibfnamefont{T.}~\bibnamefont{{Kl{\"a}hn}}},
  \bibinfo{author}{\bibfnamefont{G.}~\bibnamefont{{Mart{\'\i}nez-Pinedo}}},
  \bibinfo{author}{\bibfnamefont{W.~G.} \bibnamefont{{Newton}}},
  \bibinfo{author}{\bibfnamefont{G.}~\bibnamefont{{R{\"o}pke}}},
  \bibnamefont{and} \bibinfo{author}{\bibfnamefont{S.}~\bibnamefont{{Typel}}},
  \bibinfo{journal}{Publications of the Astronomical Society of Australia}
  \textbf{\bibinfo{volume}{34}}, \bibinfo{eid}{e067} (\bibinfo{year}{2017}),
  \eprint{1711.07411}.

\bibitem[{\citenamefont{{Demorest} et~al.}(2010)\citenamefont{{Demorest},
  {Pennucci}, {Ransom}, {Roberts}, and {Hessels}}}]{Demorest+2010}
\bibinfo{author}{\bibfnamefont{P.~B.} \bibnamefont{{Demorest}}},
  \bibinfo{author}{\bibfnamefont{T.}~\bibnamefont{{Pennucci}}},
  \bibinfo{author}{\bibfnamefont{S.~M.} \bibnamefont{{Ransom}}},
  \bibinfo{author}{\bibfnamefont{M.~S.~E.} \bibnamefont{{Roberts}}},
  \bibnamefont{and} \bibinfo{author}{\bibfnamefont{J.~W.~T.}
  \bibnamefont{{Hessels}}}, \bibinfo{journal}{\nat}
  \textbf{\bibinfo{volume}{467}}, \bibinfo{pages}{1081} (\bibinfo{year}{2010}),
  \eprint{1010.5788}.

\bibitem[{\citenamefont{{Antoniadis} et~al.}(2013)\citenamefont{{Antoniadis},
  {Freire}, {Wex}, {Tauris}, {Lynch}, {van Kerkwijk}, {Kramer}, {Bassa},
  {Dhillon}, {Driebe} et~al.}}]{Antoniadis+2013}
\bibinfo{author}{\bibfnamefont{J.}~\bibnamefont{{Antoniadis}}},
  \bibinfo{author}{\bibfnamefont{P.~C.~C.} \bibnamefont{{Freire}}},
  \bibinfo{author}{\bibfnamefont{N.}~\bibnamefont{{Wex}}},
  \bibinfo{author}{\bibfnamefont{T.~M.} \bibnamefont{{Tauris}}},
  \bibinfo{author}{\bibfnamefont{R.~S.} \bibnamefont{{Lynch}}},
  \bibinfo{author}{\bibfnamefont{M.~H.} \bibnamefont{{van Kerkwijk}}},
  \bibinfo{author}{\bibfnamefont{M.}~\bibnamefont{{Kramer}}},
  \bibinfo{author}{\bibfnamefont{C.}~\bibnamefont{{Bassa}}},
  \bibinfo{author}{\bibfnamefont{V.~S.} \bibnamefont{{Dhillon}}},
  \bibinfo{author}{\bibfnamefont{T.}~\bibnamefont{{Driebe}}},
  \bibnamefont{et~al.}, \bibinfo{journal}{Science}
  \textbf{\bibinfo{volume}{340}}, \bibinfo{pages}{448} (\bibinfo{year}{2013}),
  \eprint{1304.6875}.

\bibitem[{\citenamefont{{Cromartie} et~al.}(2020)\citenamefont{{Cromartie},
  {Fonseca}, {Ransom}, {Demorest}, {Arzoumanian}, {Blumer}, {Brook}, {DeCesar},
  {Dolch}, {Ellis} et~al.}}]{Cromartie+2020}
\bibinfo{author}{\bibfnamefont{H.~T.} \bibnamefont{{Cromartie}}},
  \bibinfo{author}{\bibfnamefont{E.}~\bibnamefont{{Fonseca}}},
  \bibinfo{author}{\bibfnamefont{S.~M.} \bibnamefont{{Ransom}}},
  \bibinfo{author}{\bibfnamefont{P.~B.} \bibnamefont{{Demorest}}},
  \bibinfo{author}{\bibfnamefont{Z.}~\bibnamefont{{Arzoumanian}}},
  \bibinfo{author}{\bibfnamefont{H.}~\bibnamefont{{Blumer}}},
  \bibinfo{author}{\bibfnamefont{P.~R.} \bibnamefont{{Brook}}},
  \bibinfo{author}{\bibfnamefont{M.~E.} \bibnamefont{{DeCesar}}},
  \bibinfo{author}{\bibfnamefont{T.}~\bibnamefont{{Dolch}}},
  \bibinfo{author}{\bibfnamefont{J.~A.} \bibnamefont{{Ellis}}},
  \bibnamefont{et~al.}, \bibinfo{journal}{Nature Astronomy}
  \textbf{\bibinfo{volume}{4}}, \bibinfo{pages}{72} (\bibinfo{year}{2020}),
  \eprint{1904.06759}.

\bibitem[{\citenamefont{{Abbott} et~al.}(2018)\citenamefont{{Abbott}, {et al.},
  {LIGO Scientific Collaboration}, and {Virgo Collaboration}}}]{Abbott+2018}
\bibinfo{author}{\bibfnamefont{B.~P.} \bibnamefont{{Abbott}}},
  \bibinfo{author}{\bibnamefont{{et al.}}}, \bibinfo{author}{\bibnamefont{{LIGO
  Scientific Collaboration}}}, \bibnamefont{and}
  \bibinfo{author}{\bibnamefont{{Virgo Collaboration}}},
  \bibinfo{journal}{\prl} \textbf{\bibinfo{volume}{121}}, \bibinfo{eid}{161101}
  (\bibinfo{year}{2018}), \eprint{1805.11581}.

\bibitem[{\citenamefont{{Bauswein} et~al.}(2017)\citenamefont{{Bauswein},
  {Just}, {Janka}, and {Stergioulas}}}]{Bauswein+2017}
\bibinfo{author}{\bibfnamefont{A.}~\bibnamefont{{Bauswein}}},
  \bibinfo{author}{\bibfnamefont{O.}~\bibnamefont{{Just}}},
  \bibinfo{author}{\bibfnamefont{H.-T.} \bibnamefont{{Janka}}},
  \bibnamefont{and}
  \bibinfo{author}{\bibfnamefont{N.}~\bibnamefont{{Stergioulas}}},
  \bibinfo{journal}{The Astrophysical Journal Letters}
  \textbf{\bibinfo{volume}{850}}, \bibinfo{eid}{L34} (\bibinfo{year}{2017}),
  \eprint{1710.06843}.

\bibitem[{\citenamefont{Essick et~al.}(2020)\citenamefont{Essick, Tews, Landry,
  Reddy, and Holz}}]{Essick:2020flb}
\bibinfo{author}{\bibfnamefont{R.}~\bibnamefont{Essick}},
  \bibinfo{author}{\bibfnamefont{I.}~\bibnamefont{Tews}},
  \bibinfo{author}{\bibfnamefont{P.}~\bibnamefont{Landry}},
  \bibinfo{author}{\bibfnamefont{S.}~\bibnamefont{Reddy}}, \bibnamefont{and}
  \bibinfo{author}{\bibfnamefont{D.~E.} \bibnamefont{Holz}}
  (\bibinfo{year}{2020}), \eprint{2004.07744}.

\bibitem[{\citenamefont{{Rampp} and {Janka}}(2002)}]{RamppJanka2002}
\bibinfo{author}{\bibfnamefont{M.}~\bibnamefont{{Rampp}}} \bibnamefont{and}
  \bibinfo{author}{\bibfnamefont{H.~T.} \bibnamefont{{Janka}}},
  \bibinfo{journal}{Astronomy \& Astrophysics} \textbf{\bibinfo{volume}{396}},
  \bibinfo{pages}{361} (\bibinfo{year}{2002}), \eprint{astro-ph/0203101}.

\bibitem[{\citenamefont{{Janka}}(2012)}]{Janka2012}
\bibinfo{author}{\bibfnamefont{H.-T.} \bibnamefont{{Janka}}},
  \bibinfo{journal}{Annual Review of Nuclear and Particle Science}
  \textbf{\bibinfo{volume}{62}}, \bibinfo{pages}{407} (\bibinfo{year}{2012}),
  \eprint{1206.2503}.

\bibitem[{\citenamefont{{Mirizzi} et~al.}(2016)\citenamefont{{Mirizzi},
  {Tamborra}, {Janka}, {Saviano}, {Scholberg}, {Bollig}, {H{\"u}depohl}, and
  {Chakraborty}}}]{Mirizzi+2016}
\bibinfo{author}{\bibfnamefont{A.}~\bibnamefont{{Mirizzi}}},
  \bibinfo{author}{\bibfnamefont{I.}~\bibnamefont{{Tamborra}}},
  \bibinfo{author}{\bibfnamefont{H.~T.} \bibnamefont{{Janka}}},
  \bibinfo{author}{\bibfnamefont{N.}~\bibnamefont{{Saviano}}},
  \bibinfo{author}{\bibfnamefont{K.}~\bibnamefont{{Scholberg}}},
  \bibinfo{author}{\bibfnamefont{R.}~\bibnamefont{{Bollig}}},
  \bibinfo{author}{\bibfnamefont{L.}~\bibnamefont{{H{\"u}depohl}}},
  \bibnamefont{and}
  \bibinfo{author}{\bibfnamefont{S.}~\bibnamefont{{Chakraborty}}},
  \bibinfo{journal}{Nuovo Cimento Rivista Serie} \textbf{\bibinfo{volume}{39}},
  \bibinfo{pages}{1} (\bibinfo{year}{2016}), \eprint{1508.00785}.

\bibitem[{\citenamefont{Steiner et~al.}(2013)\citenamefont{Steiner, Hempel, and
  Fischer}}]{Steiner:2012rk}
\bibinfo{author}{\bibfnamefont{A.~W.} \bibnamefont{Steiner}},
  \bibinfo{author}{\bibfnamefont{M.}~\bibnamefont{Hempel}}, \bibnamefont{and}
  \bibinfo{author}{\bibfnamefont{T.}~\bibnamefont{Fischer}},
  \bibinfo{journal}{Astrophys. J.} \textbf{\bibinfo{volume}{774}},
  \bibinfo{pages}{17} (\bibinfo{year}{2013}), \eprint{1207.2184}.

\bibitem[{\citenamefont{{Sukhbold} et~al.}(2018)\citenamefont{{Sukhbold},
  {Woosley}, and {Heger}}}]{Sukhbold+2018}
\bibinfo{author}{\bibfnamefont{T.}~\bibnamefont{{Sukhbold}}},
  \bibinfo{author}{\bibfnamefont{S.~E.} \bibnamefont{{Woosley}}},
  \bibnamefont{and} \bibinfo{author}{\bibfnamefont{A.}~\bibnamefont{{Heger}}},
  \bibinfo{journal}{\apj} \textbf{\bibinfo{volume}{860}}, \bibinfo{eid}{93}
  (\bibinfo{year}{2018}), \eprint{1710.03243}.

\bibitem[{\citenamefont{{Woosley} et~al.}(2002)\citenamefont{{Woosley},
  {Heger}, and {Weaver}}}]{Woosley+2002}
\bibinfo{author}{\bibfnamefont{S.~E.} \bibnamefont{{Woosley}}},
  \bibinfo{author}{\bibfnamefont{A.}~\bibnamefont{{Heger}}}, \bibnamefont{and}
  \bibinfo{author}{\bibfnamefont{T.~A.} \bibnamefont{{Weaver}}},
  \bibinfo{journal}{Reviews of Modern Physics} \textbf{\bibinfo{volume}{74}},
  \bibinfo{pages}{1015} (\bibinfo{year}{2002}),
  \urlprefix\url{http://adsabs.harvard.edu/cgi-bin/nph-bib_query?bibcode=2002RvMP...74.1015W&db_key=AST}.

\bibitem[{\citenamefont{{Woosley} and {Heger}}(2007)}]{WoosleyHeger2007}
\bibinfo{author}{\bibfnamefont{S.~E.} \bibnamefont{{Woosley}}}
  \bibnamefont{and} \bibinfo{author}{\bibfnamefont{A.}~\bibnamefont{{Heger}}},
  \bibinfo{journal}{Phys. Reports} \textbf{\bibinfo{volume}{442}},
  \bibinfo{pages}{269} (\bibinfo{year}{2007}), \eprint{astro-ph/0702176}.

\bibitem[{\citenamefont{Lattimer and Swesty}(1991)}]{Lattimer:1991nc}
\bibinfo{author}{\bibfnamefont{J.~M.} \bibnamefont{Lattimer}} \bibnamefont{and}
  \bibinfo{author}{\bibfnamefont{F.}~\bibnamefont{Swesty}},
  \bibinfo{journal}{Nucl.\ Phys.\ A} \textbf{\bibinfo{volume}{535}},
  \bibinfo{pages}{331} (\bibinfo{year}{1991}).

\bibitem[{\citenamefont{Chang et~al.}(2017)\citenamefont{Chang, Essig, and
  McDermott}}]{Chang:2016ntp}
\bibinfo{author}{\bibfnamefont{J.~H.} \bibnamefont{Chang}},
  \bibinfo{author}{\bibfnamefont{R.}~\bibnamefont{Essig}}, \bibnamefont{and}
  \bibinfo{author}{\bibfnamefont{S.~D.} \bibnamefont{McDermott}},
  \bibinfo{journal}{JHEP} \textbf{\bibinfo{volume}{01}}, \bibinfo{pages}{107}
  (\bibinfo{year}{2017}), \eprint{1611.03864}.

\bibitem[{\citenamefont{Fischer et~al.}(2016)\citenamefont{Fischer,
  Chakraborty, Giannotti, Mirizzi, Payez, and Ringwald}}]{Fischer:2016cyd}
\bibinfo{author}{\bibfnamefont{T.}~\bibnamefont{Fischer}},
  \bibinfo{author}{\bibfnamefont{S.}~\bibnamefont{Chakraborty}},
  \bibinfo{author}{\bibfnamefont{M.}~\bibnamefont{Giannotti}},
  \bibinfo{author}{\bibfnamefont{A.}~\bibnamefont{Mirizzi}},
  \bibinfo{author}{\bibfnamefont{A.}~\bibnamefont{Payez}}, \bibnamefont{and}
  \bibinfo{author}{\bibfnamefont{A.}~\bibnamefont{Ringwald}},
  \bibinfo{journal}{Phys. Rev.} \textbf{\bibinfo{volume}{D94}},
  \bibinfo{pages}{085012} (\bibinfo{year}{2016}), \eprint{1605.08780}.

\bibitem[{\citenamefont{Nakazato et~al.}(2013)\citenamefont{Nakazato,
  Sumiyoshi, Suzuki, Totani, Umeda, and Yamada}}]{Nakazato:2012qf}
\bibinfo{author}{\bibfnamefont{K.}~\bibnamefont{Nakazato}},
  \bibinfo{author}{\bibfnamefont{K.}~\bibnamefont{Sumiyoshi}},
  \bibinfo{author}{\bibfnamefont{H.}~\bibnamefont{Suzuki}},
  \bibinfo{author}{\bibfnamefont{T.}~\bibnamefont{Totani}},
  \bibinfo{author}{\bibfnamefont{H.}~\bibnamefont{Umeda}}, \bibnamefont{and}
  \bibinfo{author}{\bibfnamefont{S.}~\bibnamefont{Yamada}},
  \bibinfo{journal}{Astrophys. J. Suppl.} \textbf{\bibinfo{volume}{205}},
  \bibinfo{pages}{2} (\bibinfo{year}{2013}), \eprint{1210.6841}.

\bibitem[{\citenamefont{Raffelt}(1986)}]{PhysRevD.33.897}
\bibinfo{author}{\bibfnamefont{G.~G.} \bibnamefont{Raffelt}},
  \bibinfo{journal}{Phys. Rev. D} \textbf{\bibinfo{volume}{33}},
  \bibinfo{pages}{897} (\bibinfo{year}{1986}),
  \urlprefix\url{https://link.aps.org/doi/10.1103/PhysRevD.33.897}.

\bibitem[{\citenamefont{Payez et~al.}(2015)\citenamefont{Payez, Evoli, Fischer,
  Giannotti, Mirizzi, and Ringwald}}]{Payez:2014xsa}
\bibinfo{author}{\bibfnamefont{A.}~\bibnamefont{Payez}},
  \bibinfo{author}{\bibfnamefont{C.}~\bibnamefont{Evoli}},
  \bibinfo{author}{\bibfnamefont{T.}~\bibnamefont{Fischer}},
  \bibinfo{author}{\bibfnamefont{M.}~\bibnamefont{Giannotti}},
  \bibinfo{author}{\bibfnamefont{A.}~\bibnamefont{Mirizzi}}, \bibnamefont{and}
  \bibinfo{author}{\bibfnamefont{A.}~\bibnamefont{Ringwald}},
  \bibinfo{journal}{JCAP} \textbf{\bibinfo{volume}{1502}}, \bibinfo{pages}{006}
  (\bibinfo{year}{2015}), \eprint{1410.3747}.

\bibitem[{\citenamefont{Redondo}(2013)}]{Redondo:2013wwa}
\bibinfo{author}{\bibfnamefont{J.}~\bibnamefont{Redondo}},
  \bibinfo{journal}{JCAP} \textbf{\bibinfo{volume}{1312}}, \bibinfo{pages}{008}
  (\bibinfo{year}{2013}), \eprint{1310.0823}.

\bibitem[{\citenamefont{Burrows et~al.}(1990)\citenamefont{Burrows, Ressell,
  and Turner}}]{PhysRevD.42.3297}
\bibinfo{author}{\bibfnamefont{A.}~\bibnamefont{Burrows}},
  \bibinfo{author}{\bibfnamefont{M.~T.} \bibnamefont{Ressell}},
  \bibnamefont{and} \bibinfo{author}{\bibfnamefont{M.~S.}
  \bibnamefont{Turner}}, \bibinfo{journal}{Phys. Rev. D}
  \textbf{\bibinfo{volume}{42}}, \bibinfo{pages}{3297} (\bibinfo{year}{1990}),
  \urlprefix\url{https://link.aps.org/doi/10.1103/PhysRevD.42.3297}.

\bibitem[{\citenamefont{Croon et~al.}(2020)\citenamefont{Croon, Elor, Leane,
  and McDermott}}]{sam_muon}
\bibinfo{author}{\bibfnamefont{D.}~\bibnamefont{Croon}},
  \bibinfo{author}{\bibfnamefont{G.}~\bibnamefont{Elor}},
  \bibinfo{author}{\bibfnamefont{R.~K.} \bibnamefont{Leane}}, \bibnamefont{and}
  \bibinfo{author}{\bibfnamefont{S.~D.} \bibnamefont{McDermott}}
  (\bibinfo{year}{2020}), \eprint{2006.13942}.

\bibitem[{\citenamefont{Green}(2020)}]{Priv_Comm}
\bibinfo{author}{\bibfnamefont{D.}~\bibnamefont{Green}},
  \bibinfo{howpublished}{{private communication}} (\bibinfo{year}{2020}).

\bibitem[{\citenamefont{Aghanim et~al.}(2018)}]{Aghanim:2018eyx}
\bibinfo{author}{\bibfnamefont{N.}~\bibnamefont{Aghanim}} \bibnamefont{et~al.}
  (\bibinfo{collaboration}{Planck}) (\bibinfo{year}{2018}),
  \eprint{1807.06209}.

\end{thebibliography}

\end{document}